\documentclass{amsart} 
\usepackage{graphicx}
\usepackage{sidecap}
\usepackage{amssymb, amsmath}
\usepackage{amsfonts}
\usepackage{amssymb}
\usepackage{float}

\def\la{\label}
\def\tf{\text{ for }}

\def\bt{\begin{thm}}
\def\et{\end{thm}}

\def\bl{\begin{lem}}
\def\el{\end{lem}}

\def\bd{\begin{defi}}
\def\ed{\end{defi}}

\def\bc{\begin{cor}}
\def\ec{\end{cor}}

\def\bp{\begin{proof}}
\def\ep{\end{proof}}

\def\br{\begin{rem}}
\def\er{\end{rem}}

\newtheorem{thm}{Theorem}[section]
\newtheorem{lem}{Lemma}[section]
\newtheorem{defi}{Definition}[section]

\newtheorem{rem}{Remark}[section]
\newtheorem{cor}{Corollary}[section]

\numberwithin{equation}{section}
\numberwithin{theorem}{section}
\numberwithin{example}{section}

\numberwithin{figure}{section}

\begin{document}

\title[ENSO as Sporadic Oscillations]{El Ni\~no Southern Oscillation as Sporadic Oscillations between Metastable States}
\author[Ma]{Tian Ma}
\address[TM]{Department of Mathematics, Sichuan University,
Chengdu, P. R. China}

\author[Wang]{Shouhong Wang}
\address[SW]{Department of Mathematics,
Indiana University, Bloomington, IN 47405}
\email{showang@indiana.edu}

\thanks{The authors are grateful to Joe Tribbia for his insightful comments and for his encouragement for writing up this paper, and to Mickael Chekroun for his careful reading of the first draft of the paper. The work was supported in part by the
Office of Naval Research and by the National Science Foundation.}


\begin{abstract} The main objective of this article is to establish a new mechanism of the El Ni\~no Southern Oscillation (ENSO), as a self-organizing and self-excitation system, with  two highly coupled processes. The first is the oscillation between the two metastable  warm (El Ni\~no phase)  and cold events (La Ni\~na phase), and the second is the spatiotemporal oscillation of the  sea surface temperature (SST) field. 
The interplay between these two processes gives rises the climate variability associated with the ENSO, leads to both the random and deterministic features of the ENSO,  and 
defines a new natural {\it feedback} mechanism, which drives the sporadic oscillation of the ENSO.
The new mechanism is rigorously  derived using a dynamic transition theory developed recently by the authors, which has also been successfully applied to a wide range of problems in nonlinear  sciences. 
\end{abstract}

\maketitle
\section{Introduction}
\label{sc1}
This article  is  part of a research program  initiated recently by the authors   on dynamic  transitions  in geophysical fluid dynamics  and climate dynamic. The  main  objective  is to study the interannual low frequency variability of the atmospheric and oceanic flows, associated with typical  sources of the climate variability, including  the wind-driven (horizontal) and the thermohaline (vertical) circulations (THC) of the ocean, and the El Ni\~no Southern Oscillation (ENSO).  Their variability, independently and interactively,  may play a significant role  in climate changes, past and future. 

ENSO is the known strongest interannual climate variability associated with strong atmosphere-ocean coupling, which has significant impacts on global climate. ENSO is in fact a phenomenon that warm events (El Ni\~no phase) and clod events (La Ni\~na phase) in the equatorial eastern Pacific SST anomaly, which are associated with persistent weakening or strengthening in the trade winds.  

It is convenient and understandable to employ  simplified coupled dynamical models to investigate some essential behaviors of ENSO dynamics; see among many others \cite{SS, BH,jin,ZC,PF,JNG, neelin,neelin90,neelin91,ghil00, tribbia,penland, samelson08}.


An interesting current debate is whether ENSO is best modeled as a stochastic or chaotic system - linear and noise-forced, or nonlinear oscillatory and unstable system \cite{PF}?  It is obvious that a careful fundamental level examination of the problem is crucial.  The main objective of this article is to address this fundamental question. 

By establishing a rigorous mathematical  theory on the formation of the Walker circulation over the tropics, we  present in this article a new mechanism of the ENSO. Here we present a brief account of this new mechanism, and refer the readers to Section 5 for more precise description. 

We show that interannual variability of ENSO is the interplay of  two oscillation processes. The first is the oscillation between the metastable  warm event (El Ni\~no phase), normal event,  and cold event (La Ni\~na phase). 
We show that each metastable state has a basin of attraction, and the uncertainty of the initial states between these basins of attraction gives rise of the oscillation between these metastable states.
The  second is  the spatiotemporal oscillation of the  sea surface temperature (SST) field, which is mainly caused by the solar heating and the oceanic upwelling near Peru. 

The interplay between these two processes give rises the inter-annual variability associated with the ENSO. On the one hand, 
the metastable states oscillation has direct  influence on the SST, leading to 1) the intensification or weakening of  the oceanic upwelling, and 2) the variation of the  atmospheric Rayleigh number. 
Consequently, forcing the system  to adjust to a different metastable phase, and to intensify/weaken the corresponding event. On the other hand, the oscillation of the SST has a direct influence on the atmospheric Rayleigh number, leading to a different El Ni\~o, La Ni\~na and normal conditions.

This interplay
The above mechanism of ENSO as an interplay of the two processes defines a new natural {\it feedback} mechanism, which drives the sporadic oscillation of the ENSO, and  leads to both the random and deterministic features of the ENSO. The uncertainty  is closely related to the fluctuations between the narrow basins of attractions of the corresponding metastable events, and the deterministic feature is represented by the deterministic modeling predicting the basins of attraction and the SST. Hence  ENSO  can be considered as  a self-organizing and self-excitation system.

\medskip

The main technical method is the dynamical transition theory developed recently by the authors. The main   philosophy of the dynamic transition theory   is to search for  the full set of  transition states, giving a complete characterization on stability and  transition. The set of transition states --physical "reality" -- is represented by a local attractor. Following this philosophy, the  theory is developed  to identify the transition states and to classify them both dynamically and physically; see \cite{MW08a,MW08g} from application point of view of the theory.

 With this theory, many longstanding phase transition problems are either solved or become more accessible.  The modeling and the analysis with the applications of the theory to specific physical problems, on the one hand, provide verifications of existing experimental and theoretical studies, and, on the other hand, lead to  various new physical predictions. For example, our study  \cite{MW08f, MW08g, MW08h} of phase transitions of both liquid helium-3 and helium-4  leads not only to a theoretical understanding of the phase transitions to superfluidity observed by experiments, but also to such  physical predictions as the  existence of  a new superfluid phase C for liquid helium-3. Although these predictions need yet to be verified experimentally, they certainly  offer new insights to both theoretical and experimental studies for a better understanding of the underlying physical problems.


\section{Atmospheric Model over the Tropics}

Upwelling and zonal circulation over the tropics contains six convective cells, as shown  in Figure
\ref{f10.2}. Usually the Walker circulation is referred to the Pacific
cell over the equator, responsible for creating ocean upwelling off
the coasts of Peru and Ecuador. It was discovered by Jacob Bjerknes
in 1969 and was named after the English physicist Gilbert Walker, an
early-20th-century director of British observatories in Indian, who
discovered an indisputable link between periodic pressure variations
in the Indian Ocean and the Pacific, which he termed the "Southern
Oscillation".
\begin{figure}
  \centering
 \includegraphics[width=0.5\textwidth]{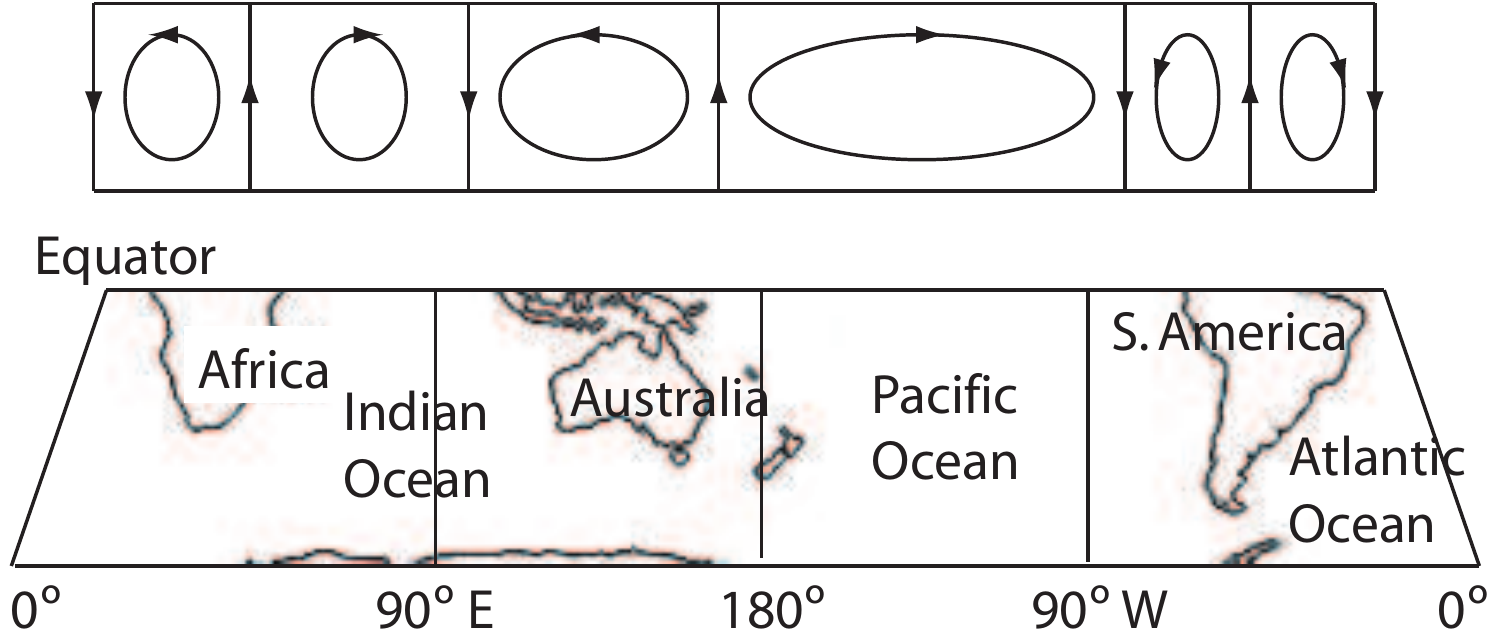}
  \caption{Global atmospheric Walker circulation over the tropics.}\la{f10.2}
 \end{figure}

The Walker cell is of such importance that its behavior is a key
factor giving rise to the El Ni\~no (more precisely, the 
El Ni\~no-Southern Oscillation (ENSO)) phenomenon. When the
convective activity weakens or reverses, an El Ni\~no
phase takes place, causing the ocean surface to be warmer than average, reducing or terminating the 
upwelling of cold water. A particularly 
strong Walker circulation causes a La Ni\~na event, resulting in
cooler ocean temperature due to stronger  upwelling; see Figure
\ref{f10.3}.
\begin{figure}  \includegraphics[width=0.6\textwidth]{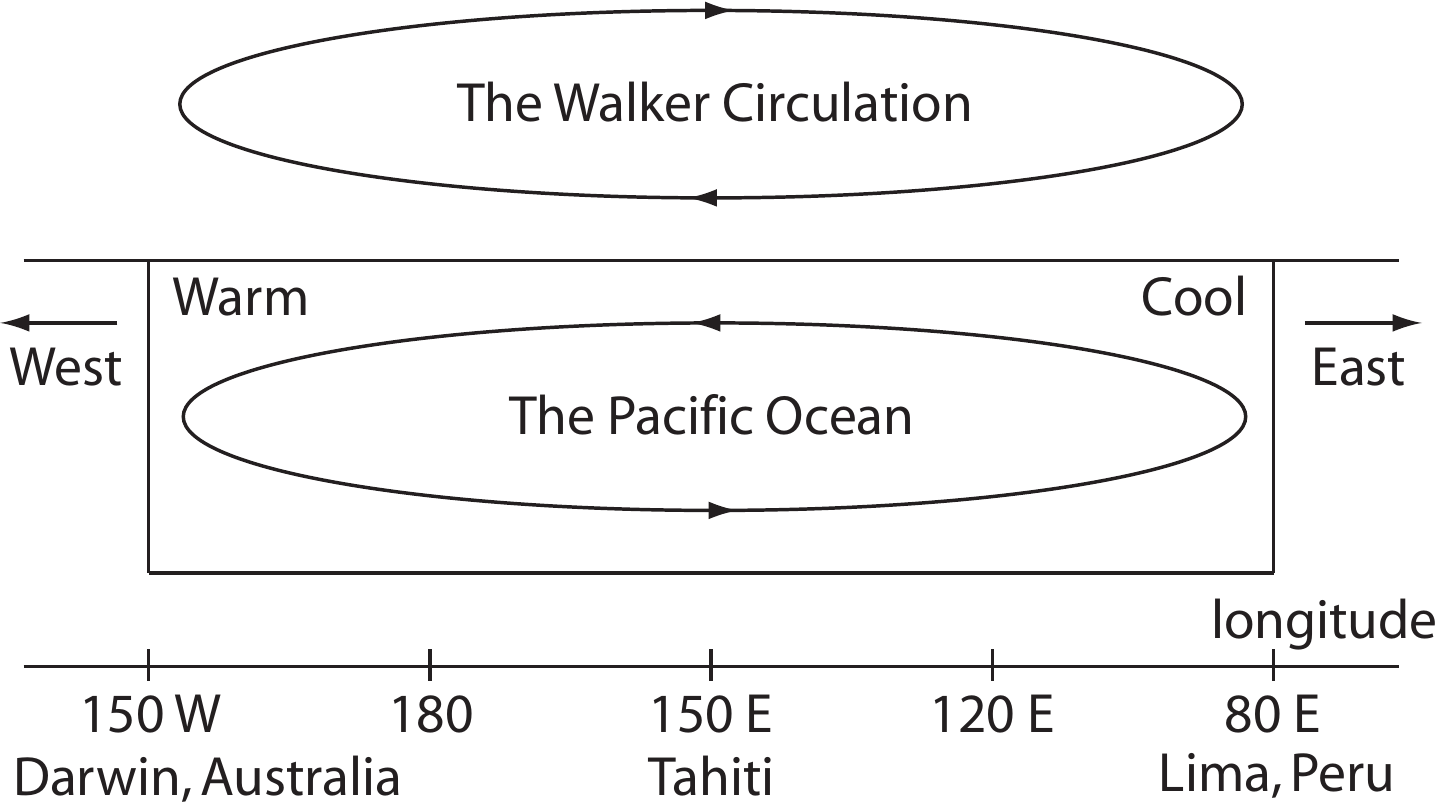}
  \caption{A schematic diagram of the Walker circulation at the
equatorial Pacific Ocean,  a main factor of the
El Ni\~no and the la Ni\~na phenomena.}\la{f10.3}
 \end{figure}
 
The atmospheric motion associated with the Walker circulation affects the loops on
either side. Under normal years, the weather behaves as expected.
But every few years, the winters becomes unusually warm or unusually
cold, or the frequency of hurricanes increases or decreases. This
entirely ocean-based cell is seen at the lower surface as easterly
trade winds  which drives the seawater and air warmed by sun moving towards the west.
The western side of the equatorial Pacific is characterized by warm,
wet low pressure weather, and the eastern side is characterized by
cool, dry high pressure weather. The ocean is about 60 cm higher in
the western Pacific as the result of this circulation. The air is
returned at the upper surface to the east, and it now becomes much
cooler and drier. An El Ni\~no phase is characterized by a
breakdown of this water and air cycle, resulting in relatively warm
water and moist air in the eastern Pacific. Meanwhile, a La
Ni\~na phase is characterized by the strengthening  of this cycle,
resulting in much cooler water and drier air than the normal
phase. During an El Ni\~no or La Ni\~na event, 
which occurs about every 3-6 years, the weather sets in for an
indeterminate period.

It is then clear that the ENSO is a coupled atmospheric-ocean phenomena. The main objective of this section is to address the mechanism of atmospheric Walker circulations over the tropics from the dynamic transition point of view. 

The atmospheric motion equations over the tropics  are the Boussinesq equations restricted on
$\theta =0$, where the meridional  velocity component
$u_{\theta}$ is set to zero, and the effect of the  turbulent friction is taking into considering using the scaling law derived in \cite{salby}:
\begin{equation}
\begin{aligned}
&\frac{\partial u_{\varphi}}{\partial t}=-(u\cdot\nabla
)u_{\varphi}-\frac{u_{\varphi}u_z}{a}+\nu\left(\Delta
u_{\varphi}+\frac{2}{a^2 }\frac{\partial
u_z}{\partial\varphi}-\frac{2u_{\varphi}}{a^2}\right)\\
&\ \ \ \ -\sigma_0u_{\varphi}-2\Omega
u_z-\frac{1}{\rho_0a}\frac{\partial p}{\partial\varphi},\\
&\frac{\partial u_z}{\partial t}=-(u\cdot\nabla
)u_z+\frac{u^2_{\varphi}}{a}+\nu\left(\Delta
u_z-\frac{2}{a^2 }\frac{\partial
u_{\varphi}}{\partial\varphi}-\frac{2u_z}{a^2 }\right)\\
&\ \ \ \ -\sigma_1u_z+2\Omega
u_{\varphi}-\frac{1}{\rho_0}\frac{\partial p}{\partial
z}-(1-\alpha_T (T-T_0))g,\\
&\frac{\partial T}{\partial t}=-(u\cdot\nabla )T+\kappa\Delta T,\\
&\frac{1}{a }\frac{\partial
u_{\varphi}}{\partial\varphi}+\frac{\partial u_z}{\partial z}=0.
\end{aligned}
\label{10.38}
\end{equation}
Here  $\sigma_i=C_i h^2$ ($i=0, 1$)  represent the turbulent friction,  $a$  is the radius of the  earth, the space domain is taken as $M=S^1_a\times (a ,a +h)$  with $S^1_a$ being the one-dimensional circle with radius $a$, and 
$$
(u\cdot\nabla )=\frac{u_{\varphi}}{a }\frac{\partial}{\partial\varphi}+u_z\frac{\partial}{\partial
z},\qquad 
\Delta =\frac{1}{a^2 }\frac{\partial^2}{\partial\varphi^2}+\frac{\partial^2}{\partial
z^2}.
$$

For simplicity, we denote
$$ (x_1,x_2)=(a \varphi ,z),\qquad (u_1,u_2)=(u_{\varphi},u_z).
$$
In atmospheric physics, the temperature $T_1$ at the tropopause
$z=a +h$ is a constant. We take $T_0$ as the average on the lower
surface $z=a $. To make the nondimensional form, let
\begin{align*}
&x=hx^{\prime},  \qquad t=h^2t^{\prime}/\kappa ,\qquad u=\kappa u^{\prime}/h, \\
&T=(T_0-T_1)T^{\prime}+T_0-(T-T_0)x^{\prime}_2,\\
&p=\kappa\nu\rho_0p^{\prime}/h^2-g\rho_0(hx^{\prime}_2+\alpha_T (T_0-T_1)hx^{\prime
2}/2).
\end{align*}
Also, we define the Rayleigh number, the Prandtl number and the scaling laws by 
\begin{equation}\label{scalinglaw}
R=\frac{\alpha_T g(T_0-T_1)h^3}{\kappa \nu},  \qquad \text{\rm Pr }=\frac{\nu}{\kappa}, \qquad 
 \delta_i=C_ih^4/\nu \quad (i=0, 1).
\end{equation}
Omitting the primes, the nondimensional form of (\ref{10.38}) reads
\begin{equation}
\begin{aligned}
&\frac{\partial u_1}{\partial t}=\text{Pr }\left[\Delta
u_1+\frac{2}{r_0}\frac{\partial u_2}{\partial
x_1}-\frac{2}{r_0}u_1-\delta_0u_1-\frac{\partial p}{\partial
x_1}\right]\\
&\ \ \ \ -\omega u_2-(u\cdot\nabla )u_1-\frac{1}{r_0}u_1u_2,\\
&\frac{\partial u_2}{\partial t}=\text{Pr }\left[\Delta
u_2-\frac{2}{r_0}\frac{\partial u_1}{\partial
x_1}-\frac{2}{r_0}u_2-\delta_1u_2+RT-\frac{\partial p}{\partial
x_2}\right]\\
&\ \ \ \ +\omega u_1-(u\cdot\nabla )u_2-\frac{1}{r_0}u^2_1,\\
&\frac{\partial T}{\partial t}=\Delta T+u_2-(u\cdot\nabla )T,\\
&\frac{\partial u_1}{\partial x_1}+\frac{\partial u_2}{\partial
x_2}=0,
\end{aligned}
\label{10.39}
\end{equation}
where $(x_1,x_2)\in M=(0,2\pi r_0)\times (r_0,r_0+1)$, $\delta_0$ and
$\delta_1$ are as in (\ref{scalinglaw}), $(u\cdot\nabla )$ and $\Delta$
as usual differential operators, and
\begin{equation}
\omega =\frac{2\Omega h^2}{\kappa}.\label{10.40}
\end{equation}

The problem is supplemented with the natural periodic boundary condition in the  $x_1$-direction, and the free-slip boundary condition on the top and bottom boundary:
\begin{align}
& (u, T)(x_1+2\pi r_0,x_2, t)=(u, T)(x_1,x_2, t),  \label{10.41}\\
& 
\left\{
\begin{aligned} 
&u_2=0,\ \frac{\partial u_1}{\partial
x_2}=0,  T=\varphi (x_1) \qquad   && \text{at}\ x_2=r_0,\\
&u_2=0,\ \frac{\partial u_1}{\partial
x_2}=0,\ T=0\qquad  &&  \text{at}\ x_2=r_0+1.
\end{aligned}
\right. \label{10.44}
\end{align}
Here $\varphi (x_1)$ is the temperature deviation from the average
$T_0$ on the equatorial surface and is periodic, i.e.,
$$\int^{2\pi r_0}_0\varphi (x_1)dx_1=0\ \ \ \ \text{and  }\ \varphi
(x_1)=\varphi (x_1+2\pi r_0).$$ The
deviation $\varphi (x_1)$ is mainly caused by a difference in the
specific heat capacities between the sea water and land.

\section{Walker Circulation under the Idealized Conditions}
In an idealized case, the temperature deviation
$\varphi$ vanishes. In this case, the study of transition of
(\ref{10.39}) is of special importance to understand the
longitudinal circulation. Here, we are devoted to discuss the
dynamic bifurcation of (\ref{10.39}), the Walker cell structure of
bifurcated solutions, and the convection scale under the idealized
boundary condition
\begin{equation}
\varphi (x_1)=0 \qquad 
\text{ for any }  0\leq x_1\leq 2\pi r_0.\label{10.45}
\end{equation}

The following theorem provides a theoretical basis to  understand  
the equatorial Walker circulation. The proof of this theorem follows the same method as given in 
\cite{MW07a}, where similar results are obtained  for the classical B\'enard convection.

\bt\label{t10.1} Under the idealized condition (\ref{10.45}), the problem (\ref{10.39}) with  (\ref{10.41})   and (\ref{10.44}) undergoes  a Type-I transition \footnote{See \cite{MW08a} for  precise definition.} at the critical Rayleigh number $R=R_c$. 
More precisely, the following statements hold true:

\begin{itemize}

\item[(1)] When the Rayleigh number $R\leq R_c$, the equilibrium
solution $(u,T)=0$ is globally stable in the phase space $H$ defined by:
$$H=\{(u,T)\in L^2(M)^3\ |\ \text{\rm div} u=0, u_2|_{x_2=r_0, r_0 +1}=0, (u,T)\ \text{satisfies}\
(\ref{10.41}) \},$$
where $L^2(M)$  stands for the square-integrable functions in $M$.

\item[(2)] When $R_c<R<R_c+\delta$ for some $\delta >0$, this
problem bifurcates from $((u,T),R)=(0,R_c)$ to an attractor
${\mathcal{A}}_R=S^1$, consisting of steady state
solutions, which attracts $H \setminus \Gamma$ ,   where $\Gamma$ is the stable manifold
of $(u, T)=0$ with codimension two.

\item[(3)] For each steady state solution
$\psi_R=(u_R,T_R)\in{\mathcal{R}},u_R$ is topologically equivalent
to the structure as shown in Figure \ref{f10.4}.

\item[(4)] For any initial value $\psi_0=(u_0,T_0)\in H\setminus (\Gamma\cup
E)$, there exists a time $t_0\geq 0$ such that for any $t>t_0$ the
velocity field $u(t,\psi_0)$ is topologically equivalent to the
structure as shown in Figure \ref{f10.5} either (a) or (b), where $\psi
=(u(t,\psi_0),T(t,\psi_0))$ is the solutions of the problem with
$\psi (0)=\psi_0$, and
$$E=\{(u,T)\in H|\ \int^{r_0+1}_{r_0}u_1dx_2=0\}.$$
\end{itemize}
\et
\begin{figure}
  \centering
  \includegraphics[width=0.6\textwidth]{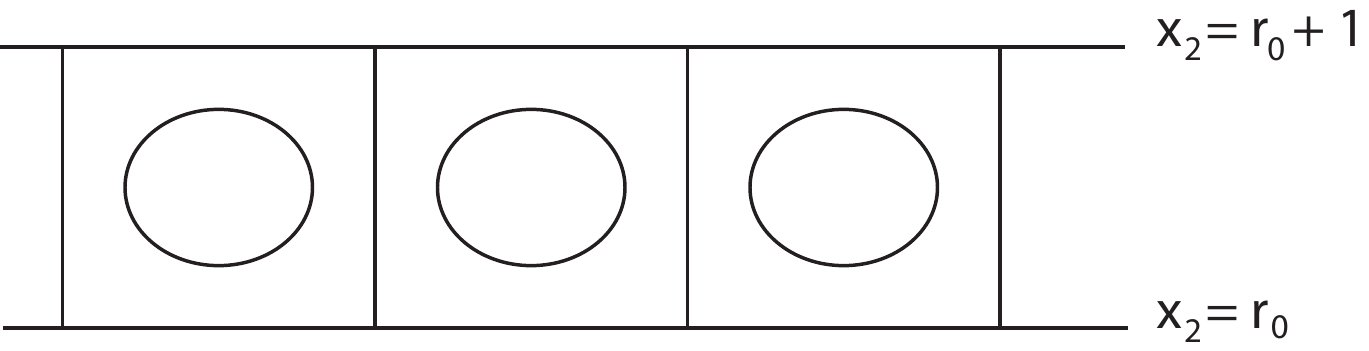}
  \caption{The cell structure of the steady state solutions in the
bifurcated attractor ${\mathcal{A}}_R$.}\la{f10.4}
 \end{figure}

\begin{figure}[hbt]
  \centering
  \includegraphics[width=0.49\textwidth]{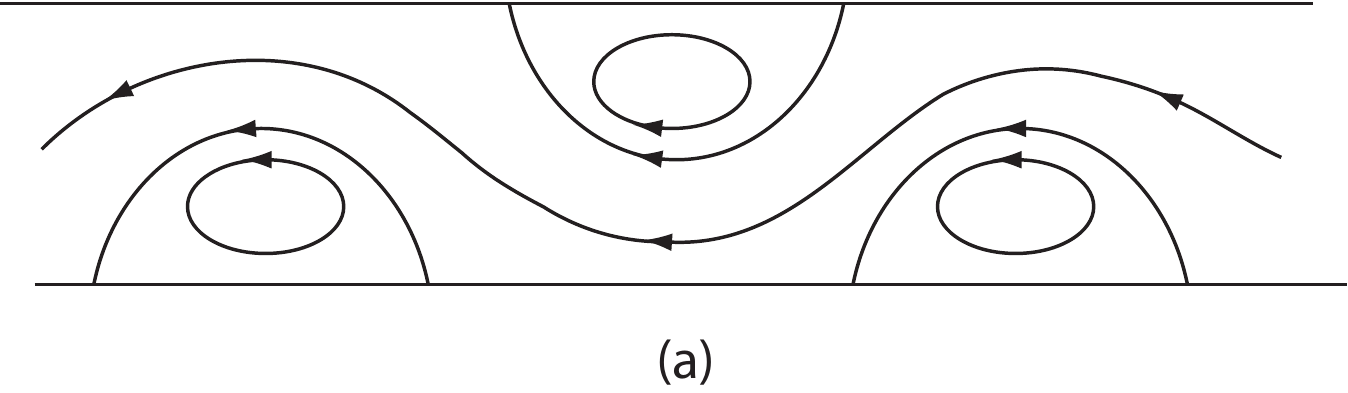} 
  \includegraphics[width=0.49\textwidth]{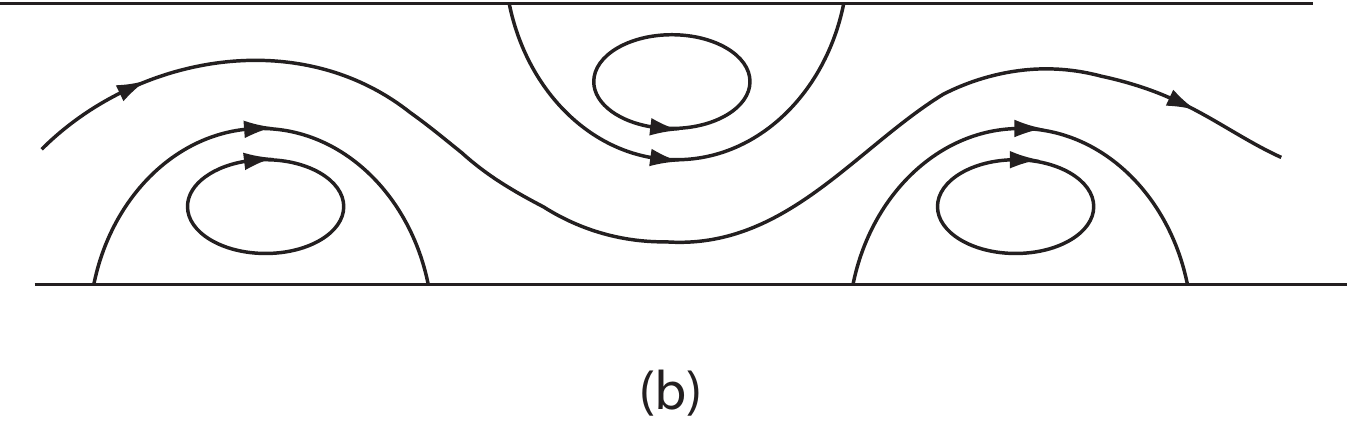}
  \caption{The Walker cell structure with the cells separated by a
cross channel flow: (a) a westbound cross channel flow, and (b) an
eastbound cross channel flow. This cross-channel  flow pattern has the same topological structure as
the Walker circulation over the tropics and the Branstator-Kushnir
waves in the atmospheric dynamics \cite{branstator,kush}.
}\la{f10.5}
 \end{figure}

A few remarks are now in order:

\medskip

{\sc First}, in a more realized case where  $T=T_0+\varphi (x_1)$ at $x_2=r_0$, 
the temperature deviation $\varphi (x_1)$ is small in
comparison with the average temperature gradient $T_0-T_1$. Therefore, the
more realistic case can be considered as a perturbation of the idealized
case.

\medskip

{\sc Second},  mathematically, the conclusion that
${\mathcal{A}}_R=S^1$ consisting of steady state solutions is
derived from the invariance of (\ref{10.39}) under a translation
$x_1\rightarrow x_1+\alpha$. It implies that for any two solutions
$\psi_1,\psi_2\in{\mathcal{A}}_R$, they are the same up to a phase
angle $\alpha$ in the longitude, namely
$$\psi_2(x_1,x_2)=\psi_1(x_1+\alpha ,x_2),\ \text{for\ some}\
\alpha\in [0,2\pi r_0].$$ 
For  the idealized case, this conclusion
is nature as   the equator is treated as homogeneous.

In addition,  Assertion (2) amounts to saying that as the initial
value $\psi_0$ varies, i.e., an external perturbation is applied,
 the rolls of the Walker circulation will be
translated by a phase angle $\alpha$.
This behavior is
termed as the translation oscillation, which can be used to explain
the ENSO phenomenon for the idealized circumstance.

\medskip

{\sc Third}, when a deviation $\varphi (x_1)$ is present, the
translation symmetry is broken, and consequently,  the bifurcated
attractor ${\mathcal{A}}_R$ consists of isolated equilibrium
solutions instead of a circle $S^1$. Thus, the mechanism of the ENSO will be explained as state
exchanges between these equilibrium solutions in ${\mathcal{A}}_R$,
which are metastable. We shall address this point in detail later.

\medskip

{\sc Fourth},  by the structural stability theorems in \cite{amsbook}, we see  from Assertion (4) that the roll structures
illustrated by Figure \ref{f10.5} (a) and (b) are structurally stable.
Hence, for a deviation perturbation $\varphi (x_1)$, these
structures are not destroyed, providing a realistic  characterization of  the
Walker circulation.

\medskip

{\sc Fifth}, in Assertion (3), these rolls in the bifurcated
solutions are closed. However, under some external perturbation, a
cross channel flow appears which separates the rolls apart, and
globally transports heat between them.

\medskip

{\sc Sixth,}   it is then classical   \cite{MW07a, b-book,chandrasekhar, dr} to show that  the 
 critical Rayleigh number  is  given by 
\begin{equation}
R_c=\min\limits_{\alpha^2}\left[(\alpha^2+\pi^2)\delta_1+\frac{(\alpha^2+\pi^2)^3+\pi^2(\alpha^2
+\pi^2)\delta_0}{\alpha^2}\right].\label{10.26}
\end{equation}
Let $g(x)=(x+\pi^2)\delta_1+((x+\pi^2)^3+\pi^2(x+\pi^2)\delta_0)/x$.
Then, $x_c=\alpha^2_c$ satisfies $g^{\prime}(x)=0$. Thus we get
\begin{equation}
\delta_1\alpha^4_c-(\alpha^2_c+\pi^2)^3-\pi^2(\alpha^2_c+\pi^2)\delta_0+3\alpha^2_c(\alpha^2_c+
\pi^2)^2+\pi^2\alpha^2_c\delta_0=0.\label{10.27}
\end{equation}
We assume that $\delta_1>\delta_0$ for $h\neq 0$. By
$\alpha^2_c=\pi^2/L^2_c$, when $h$ is large, $\delta_1\gg 1$
and $\delta_1\gg\delta_0$, from (\ref{10.26}) and (\ref{10.27}) we
derive that
\begin{align}
& \alpha^4_c\cong\frac{\pi^4(\pi^2+\delta_0)}{\delta_1},\label{10.28}\\
&
L^2_c\cong\delta^{{1}/{2}}_1/(\pi^2+\delta_0)^{{1}/{2}},\label{10.29}
\\
&
R_c\cong\pi^2\delta_1+\pi^2\delta_0L^2_c,\label{10.30}
\end{align}
These formulas (\ref{10.28})-(\ref{10.30}) are very useful in studying
large-scale convection motion, as shown in the next section.



\section{Walker circulation under natural conditions}

\subsection{Physical parameters  and effects of the turbulent friction terms}
It is known that air properties vary with the temperature and
pressure. The table below lists some common properties of air:
density,  kinematic viscosity, thermal diffusivity, expansion
coefficient and   the Prandtl number for temperatures
between-$100^{\circ}$C and $100^{\circ}$C. {\footnotesize
\begin{center}
\begin{tabular}{|c|c|c|c|c|c|}
\hline $\begin{aligned} &\text{Temperature}\\
&\ \ t\\
&(^{\circ}C)\end{aligned}$& $\begin{aligned} &\text{Density}\\
&\ \ \rho\\
&(\text{kg/m}^3)
\end{aligned}
$&$\begin{aligned} &\text{Kinematic}\\
&\text{viscosity}\\
&\ \ \nu\\
&(\text{m}^2/\text{s})\times 10^{-6}
\end{aligned}$&
$\begin{aligned} &\text{Thermal}\\
&\text{Diffusivity}\\
&\ \ \kappa\\
&(\text{m}^2/\text{s})\times 10^{-6}
\end{aligned}
$&$\begin{aligned} &\text{Expansion}\\
&\text{coefficient}\\
&\ \ a\\
&(1/\text{k})\times 10^{-3}
\end{aligned}
$&$\begin{aligned} &\text{Prandtl}\\
&\text{number}\\
&\ \ \text{Pr }
\end{aligned}
$\\
\hline -100&1.980&5.95&8.4&5.82&0.74\\
\hline -50&1.534&9.55&13.17&4.51&0.725\\
\hline 0&1.293&13.30&18.60&3.67&0.715\\
\hline 20&1.205&15.11&21.19&3.40&0.713\\
\hline 40&1.127&16.97&23.87&3.20&0.711\\
\hline 60&1.067&18.90&26.66&3.00&0.709\\
\hline 80&1.000&20.94&29.58&2.83&0.708\\
\hline 100&0.946&23.06&32.8&2.68&0.703\\
\hline
\end{tabular}
\label{table10.1}
\end{center}}

The average temperature over the equator is about $20\sim
30^{\circ}$C. Based on the data in Table \ref{table10.1}, we take a set of
physical parameters of air for the Walker circulation as follows:
\begin{eqnarray*}
&&\nu =1.6\times 10^{-5}\text{m}^2/\text{s},\ \ \ \ \kappa =2.25\times
10^{-5}\text{m}^2/\text{s},\\
&&\alpha_T=3.3\times 10^{-3} / {}^{\circ}C,\ \ \ \ \text{Pr }=0.71.
\end{eqnarray*}
The height $h$ of the troposphere is taken by
$$h=8\times 10^3\text{m}.$$
Thus, the Rayleigh number for the equatorial atmosphere is
\begin{equation}
R=\frac{g\alpha_T(T_0-T_1)}{\kappa\nu}h^3=3.6\times
10^{19}(T_0-T_1)/  {}^{\circ}C.\label{10.65}
\end{equation}

In the classical theory, we have the critical Rayleigh number
\begin{equation}
R_c=\min\limits_{\alpha^2}\frac{(\alpha^2+\pi^2)^3}{\alpha^2}=\frac{27}{4}\pi^4=675,\label{10.66}
\end{equation}
and the convective scale
\begin{equation}
L^2_c=\pi^2/\alpha^2_c=2.\label{10.67}
\end{equation}
Namely, the diameter of convective roll is
$$d=L_c\times h=20\text{km}.$$
However, based on atmospheric observations, there are six cells
 overall tropics,  and  it shows that the real convective scale is
about
\begin{equation}
L=\frac{a \pi}{3} =6600 \text{ km }. \label{10.68}
\end{equation}

As a comparison, the values (\ref{10.66}) and (\ref{10.67}) from the
classical theory are too small to match the realistic data  given in 
(\ref{10.65}) and (\ref{10.68}).

We use (\ref{10.28})-(\ref{10.30}) to discuss
this problem. With  the values  in (\ref{10.65}) and
(\ref{10.68}), we take
\begin{equation}
\delta_1=2.7\times 10^{20},\ \ \ \ \delta_0=3.5\times
10^8.\label{10.69}
\end{equation}
In comparison with (\ref{scalinglaw}), for air, the constants $C_0$ and
$C_1$ take the following values
$$ C_0=1.37\times 10^{-12}  \ m^{-2}\cdot s^{-1},\qquad C_1=1.05 \ m^{-2}\cdot s^{-1}.
$$
Then, the convective scale $L_c$ in (\ref{10.29}) takes the value
\begin{equation}
L_c=[\delta_1/\delta_0]^{{1}/{4}}h=7500\text{km},\label{10.70}
\end{equation}
which is slightly bigger than the realistic value $L\cong
6300\text{km}$. And, the critical Rayleigh number in (\ref{10.30})
takes
$$R_c\cong\pi^2\delta_1+\pi^2\delta_0L^2_c\cong 2.7\times 10^{21}.$$
Thus, the critical temperature difference is
\begin{equation}
\Delta T_c\cong 60^{\circ}C, \label{10.71}
\end{equation}
which is realistic, although it  is slightly smaller than the realistic temperature
difference on the equator.

However, with the classical theory with no friction terms,   the critical values  would be  $ L_c \cong 20 km$   and  $\Delta T_c\cong
10^{-17\circ}C$, which are certainly unrealistic.

\subsection{Transition in natural conditions}
We now return the natural boundary condition
$$\varphi (x_1)\not\equiv 0.$$ 
In this case, equations (\ref{10.39})
admits a steady state solution
\begin{equation}
\tilde{\psi}=(V,J),\ \ \ \ (V=(V_1,V_2)).\label{10.72}
\end{equation}
Consider the deviation from this basic state:
$$u\rightarrow u+V,\ \ \ \ T\rightarrow T+J.$$
Then (\ref{10.39}) becomes
\begin{equation}
\begin{aligned}
&\frac{\partial u_1}{\partial t}+(u\cdot\nabla
)u_1+\frac{u_1u_2}{r_0}=\text{Pr }[\Delta
u_1-\delta^{\prime}_0u_1-\frac{\partial p}{\partial x_1}]\\
&\ \ \ \ -(u\cdot\nabla )u_1-(u\cdot\nabla
)V_1-\frac{V_1}{r_0}u_2-\frac{V_2}{r_0}u_1,\\
&\frac{\partial u_2}{\partial t}+(u\cdot\nabla
)u_2-\frac{u^2_1}{r_0}=\text{Pr }[\Delta
u_2-\delta^{\prime}_1u_2+RT-\frac{\partial p}{\partial x_2}]\\
&\ \ \ \ -(u\cdot\nabla )u_2-(u\cdot\nabla
)V_2+\frac{2V_1}{r_0}u_1,\\
&\frac{\partial T}{\partial t}+(u\cdot\nabla )T=\Delta
T+u_2-(u\cdot\nabla )J-(u\cdot\nabla )T,\\
&\text{div} u=0.
\end{aligned}
\label{10.73}
\end{equation}
The boundary conditions are the free-free boundary conditions given by
\begin{equation}
\begin{aligned} 
&(u,T)\ \text{is\ periodic\ in}\
x_1-\text{direction},\\
&T=0,u_2=0,\frac{\partial u_1}{\partial x_2}=0\ \text{at}\
x_2=r_0,r_0+1.
\end{aligned}
\label{10.74}
\end{equation}

It is known that $|\varphi (x_1)|$, with $\Delta T=T_0-T_1\cong
100^{\circ}C$ as unit, is small. Hence, the steady state solution
$(V,J)$ is also small:
$$\|(V,J)\|_{L^2}=\varepsilon\ll 1.$$
Thus, (\ref{10.73}) is a perturbated equation of (\ref{10.39}).

Since perturbation terms involving $(V, J)$  are  not invariant under the zonal  translation
(in the $x_1$-direction), for general small functions $\varphi
(x_1)\neq 0$, the first eigenvalues of (\ref{10.73}) are (real or complex) simple, and
by the perturbation theorems in \cite{b-book}, all eigenvalues of linearized equation of (\ref{10.73}) satisfy
the  following principle of exchange of stability (PES):
\begin{align*}
&
\text{Re}\beta^{\varepsilon}_i(R)
\left\{\begin{aligned}
&<0&& \tf  R<R^{\varepsilon}_c,\\
&=0&&\tf R=R^{\varepsilon}_c,\\
&>0&&\tf  R>R^{\varepsilon}_c, 
\end{aligned} \right.   && \text{ for any } 1\leq i\leq m,\\
&  \text{Re}\beta^e_j(R^{\varepsilon}_c)<0  &&  \text{ for any } j\geq
m+1,
\end{align*}
where $m=1$ as $\beta^{\varepsilon}_1(R)$ is real, $m=2$ as
$\beta^{\varepsilon}_1(R)$ is complex near $R^{\varepsilon}_c$, and
$R^{\varepsilon}_c$ is the critical Rayleigh number of perturbed
system (\ref{10.73}).


The following two  theorems follow directly from Theorem~\ref{t10.1} and the perturbation theorems in \cite{b-book}.

\bt\la{t10.2}
Let $\beta^{\varepsilon}_1(R)$ near
$R=R^{\varepsilon}_c$ be a real eigenvalue. Then the system
(\ref{10.73}) has a transition at $R=R^{\varepsilon}_c$, which is
either mixed (Type-III) or continuous (Type-I), depending on the
temperature deviation $\varphi (x_1)$. Moreover, we have the
following assertions:

\begin{itemize}
\item[(1)] If the transition is Type-I, then as
$R^{\varepsilon}_c<R<R^{\varepsilon}_c+\delta$ for some $\delta >0$,
the system bifurcates at $R^{\varepsilon}_c$ to exactly two steady
state solutions $\psi_1$ and $\psi_2$ in $H$, which are
attractors. In particular, space $H$ can be decomposed into two open
sets $U_1,U_2$:
$$H=\bar{U}_1+\bar{U}_2,\ \ \ \ U_1\cap U_2=\phi ,\ \ \ \ \psi
=0\in\partial U_1\cap\partial U_2,$$ such that $\psi_i\in
U_i$  $(i=1,2)$, and $\psi_i$ attracts $U_i$.

\item[(2)] If the transition is Type-III, then there is a saddle-node
bifurcation at $R=R^*$ with $R^*<R^{\varepsilon}_c$ such that  the following statements hold true:

\begin{itemize}

\item[(a)] if  $R^*<R<R^{\varepsilon}_c+\delta$ with $R\neq R^{\varepsilon}_c$,
the system has two steady state solutions $\psi^R_+$ and $\psi^R_-$
which are attractors, as shown in  Figure~\ref{f10.10}, such that 
$$\psi^R_+ = 0 \qquad \text{ for } R^*<R<R^{\varepsilon}_c.$$

\item[(b)]  There is an open set $U\subset H$ with $0\in U$ which
can be decomposed into two disjoint open sets
$\bar{U}=\bar{U}^R_+ +\bar{U}^R_-$ with $\psi^R_\pm \in U^R_\pm$ and
$\psi^R_\pm$ attracts $U^R_\pm$.
\end{itemize}

\item[(3)] For any initial value $\psi_0=(u_0,T_0)\in U^R_-$ for $R > R^\ast$ or 
$\psi_0=(u_0,T_0)\in U^R_+$  for $R > R^\varepsilon_c$, there exists a time $t_0\geq 0$ such that for any $t>t_0$ the
velocity field $u(t,\psi_0)$ is topologically equivalent to the
structure as shown in Figure \ref{f10.5} either (a) or (b), where $\psi
=(u(t,\psi_0),T(t,\psi_0))$ is the solutions of the problem with
$\psi (0)=\psi_0$.
\end{itemize}
\et


\bt\la{t10.3}
Let $\beta^{\varepsilon}_1(R)$ be complex near
$R=R^{\varepsilon}_c$. Then the system (\ref{10.73}) bifurcates from
$(\psi ,R)=(0,R^{\varepsilon}_c)$ to a periodic solution $\psi_R(t)$
on $R^{\varepsilon}_c<R$, which is an attractor, and $\psi_R(t)$ can
be expressed as
\begin{equation}
\psi_R(t)=A_R(\cos\rho t\psi_1+\sin\rho
t\tilde{\psi}_1)+o(|A_R|,\varepsilon ),\label{10.77}
\end{equation}
where $A_R=\alpha (R-R^{\varepsilon}_c),\alpha$ and $\rho$ are
constants depending on $\varphi (x_1)$, and $\psi_1,\tilde{\psi}_1$
 are first eigenfunctions of linearized equations of (\ref{10.39}).
\et

A few remarks are now in order.

\medskip

{\sc First,} Theorems \ref{t10.1}, \ref{t10.2}   and  \ref{t10.3} provide the possible dynamical behaviors for the
zonal atmospheric  circulation  over the tropics.  Theorem \ref{t10.1} describes  the translation oscillation, and does not represent a realistic explanation to the ENSO.

\medskip

{\sc Second,} the periodic solution (\ref{10.77}) characterizes a roll
pattern translating with a constant velocity over the  equator,
eastward or westward, as shown in Figure \ref{f10.7} (a) and (b)
The time-periodic oscillation  obtained in Theorem \ref{t10.3}   
does not represent the typical oscillation in the ENSO phenomena, as the Walker circulation does not obey this zonal translational oscillation.
\begin{figure}[hbt]
  \centering
  \includegraphics[width=0.3\textwidth]{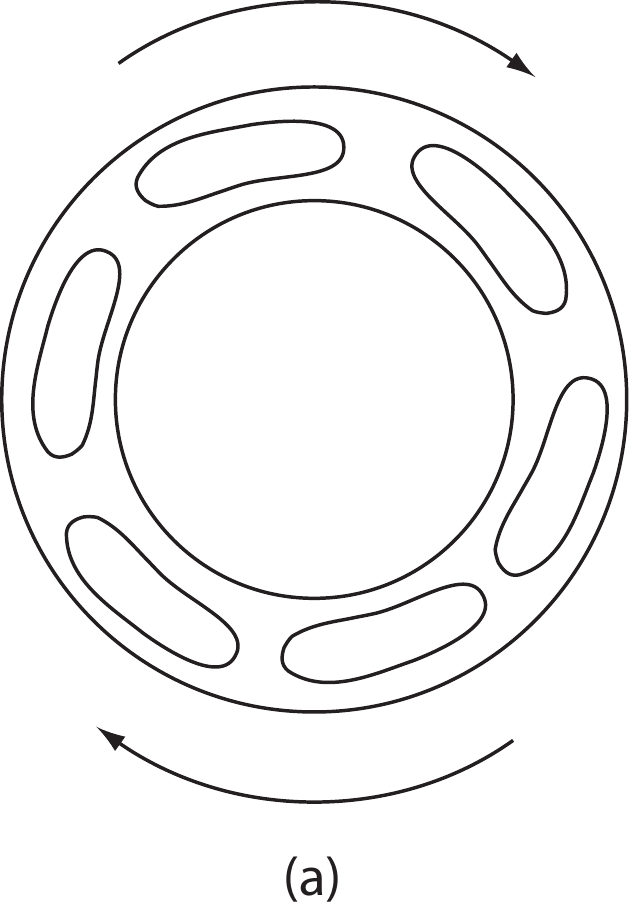} \qquad
  \includegraphics[width=0.3\textwidth]{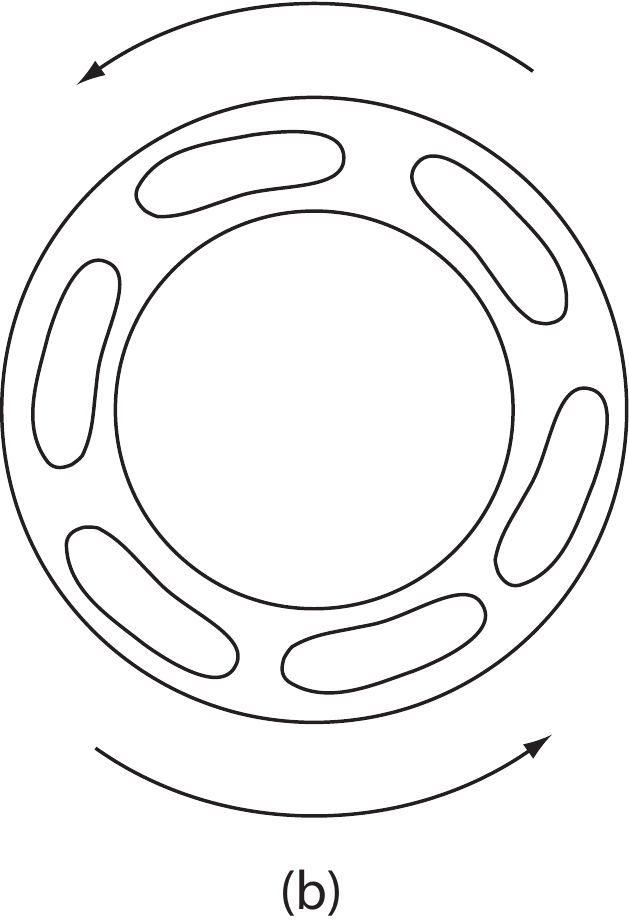}
  \caption{Time-periodic translation of the Walker circulation
pattern; (a) an eastward translation, (b) a westward translation.}\la{f10.7}
 \end{figure}

\medskip

{\sc Third},  Theorem \ref{t10.2}  characterizes the  oscillation between metastable states, 
The oscillation between the metastable states in Theorem~\ref{t10.2} can be either Type-I or Type-III, depending on the number
$$b=<G(\Psi_1),\Psi^*_1>,$$
where $\Psi_1$ is the first eigenvector of the linearized equation of (\ref{10.73}) at
$R=R^{\varepsilon}_c$,  $\Psi^*$ is the corresponding  first eigenvector of
the adjoint linearized problem, and  $G$ represents the nonlinear terms in (\ref{10.73}). 

Namely, if  $b\neq 0$,  the transition is Type-III  and if  $b=0$, the transition is Type-I.
From mathematical viewpoint, for almost all
functions $\varphi$, $b\neq 0$. Hence, the case where $b \neq 0$,  consequently 
the Type-III transition in Theorem~\ref{t10.2}, is generic. 
In other words, the  Type-III transition derived in this theorem provides a correct oscillation mechanism of the ENSO between two metastable El Ni\~no and La Ni\~na events, and we shall explore this point of view in the next section in detail.

\section{Metastable Oscillation Theory of  ENSO}
It is well known that the Walker cell at the equatorial
Pacific Ocean is closely related to the ENSO phenomenon. Its behavior is the key to the  understanding of ENSO.

\medskip

{\bf Southern Oscillation Phenomenon.}
The Southern Oscillation Index (SOI) gives a simple measure of the
strength and phase of the Southern Oscillation, and indicates the
state of the Walker circulation. The SOI is defined by the pressure
difference between Tahiti and Darwin. When the Walker circulation
enters its El Ni\~no phase, the SOI is strongly negative; 
when it is in the La Ni\~na phase, the SOI is strongly
positive; and in the normal state the SOI is small; see Figure
\ref{f10.8}.

\begin{SCfigure}[25][t]
  \centering
  \includegraphics[width=0.4\textwidth]{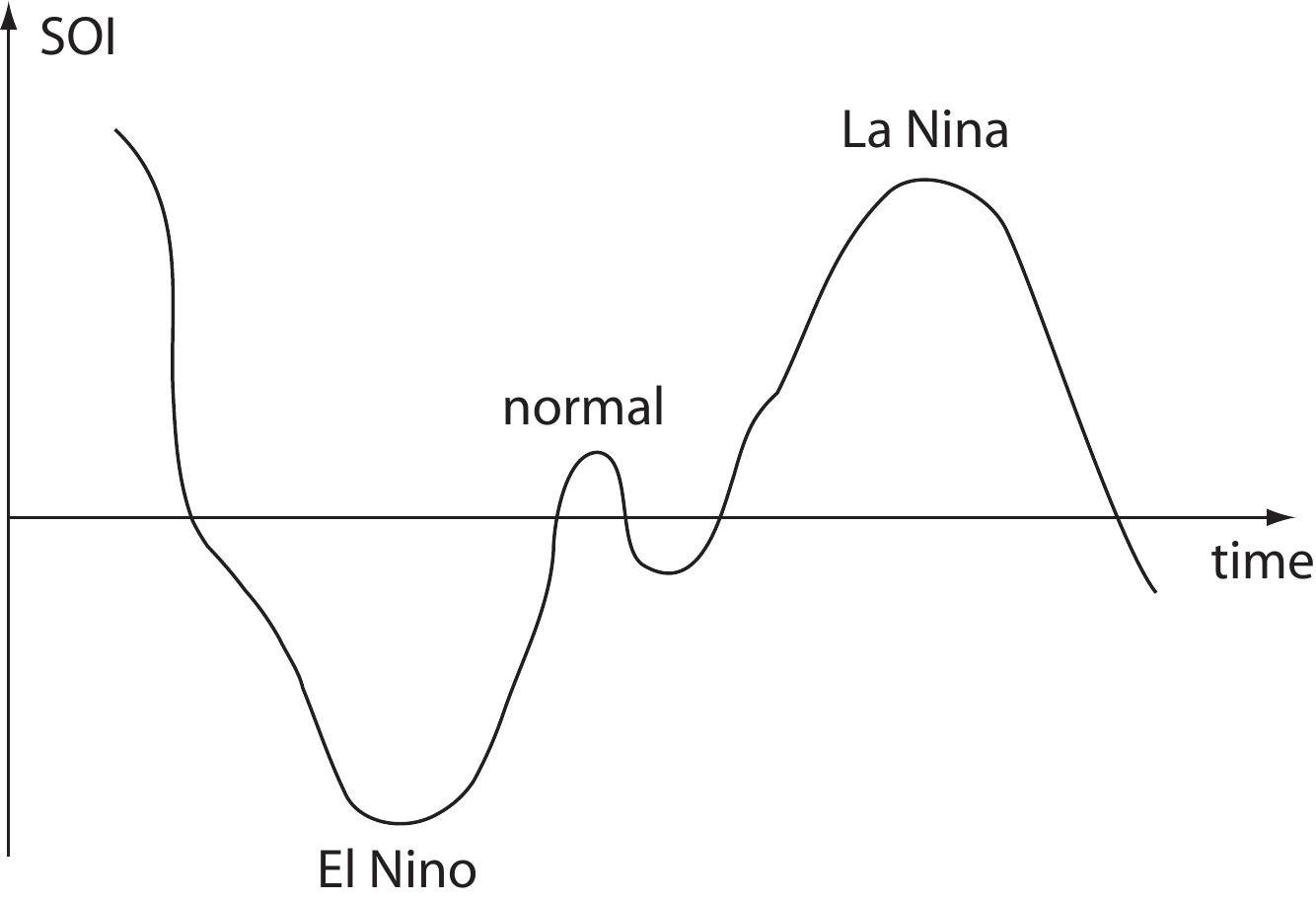}
  \caption{SOI indicates the state of the Walker circulation.}\la{f10.8}
 \end{SCfigure}
 
This point   can be further examined by further observational  data of
SOI  as shown in Figure \ref{f10.9}. In the SOI diagram, we observe that there are two groups of
oscillations: the relatively large amplitude fluctuation and the
relatively small amplitude fluctuation. The large one occurred in
1950-1951, 1955-1956, and 1974-1975 for positive SOI higher than 13,
in 1965-1966, 1972-1973, 1977-1978, 1981-1982, and 1987-1988 for
negative SOI lower than -12. The small one with SOI between-8 to 8
appeared in other years.
\begin{figure}[hbt]
  \centering
\includegraphics[width=0.95\textwidth]{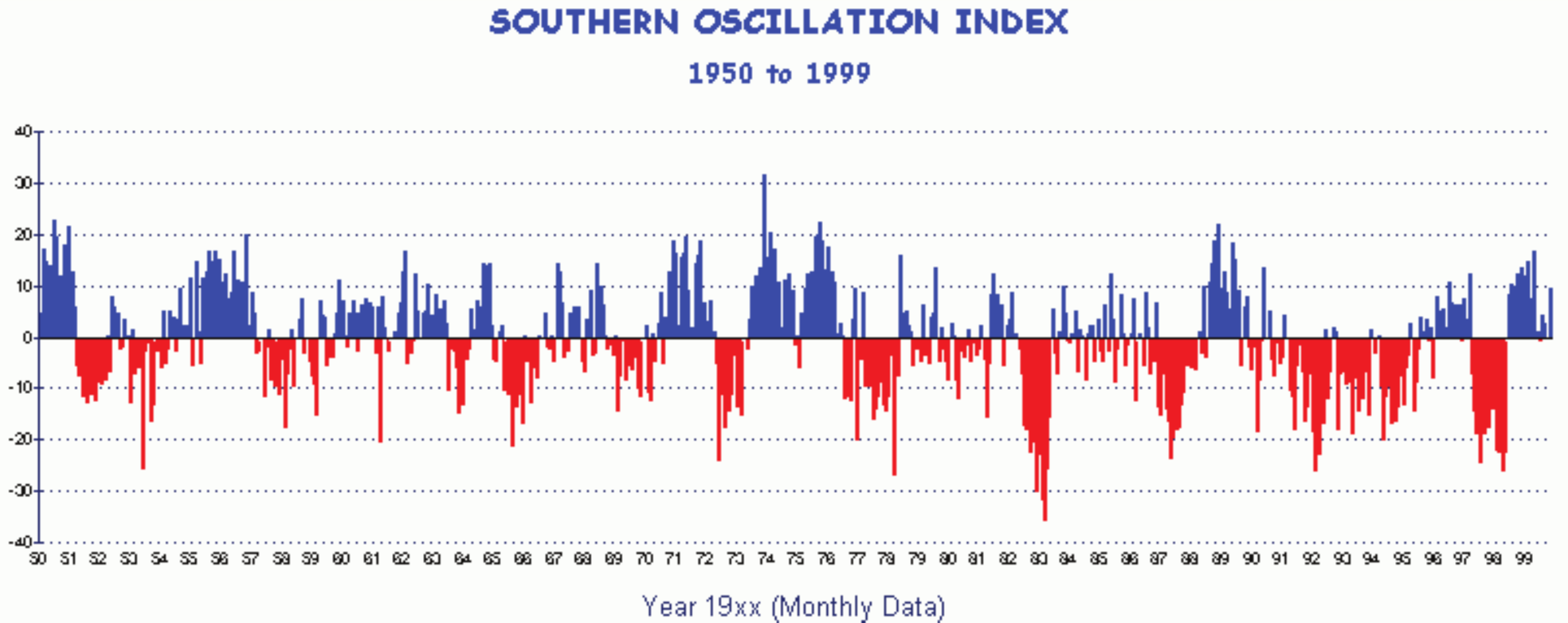}
  \caption{SOI diagram, adopted from Web: BOM-Monitoring
Climate-Climate variability and El Ni\~no.}\la{f10.9}
 \end{figure}

{\bf El Ni\~no and La Ni\~na States.} 
The non-homogeneous temperature distribution shows that  the locations of
the cells over the tropics are relatively fixed. 
This suggests that the homogeneous case as described by Theorem \ref{t10.1} is not a valid description of the realistic situation. In addition,  the translation oscillation formation, suggest in Theorem~\ref{t10.3},  does not describe the  ENSO either. 

Hence the nature theory for the atmospheric circulation over the tropics associated with the Walker circulation and the ENSO is  given by Theorem~\ref{t10.2}.  It is
known that the steady state solutions of (\ref{10.39}) can be
written as 
\begin{equation} 
\tilde{\psi}+\psi_R,\label{10.78}
\end{equation}
where $\tilde{\psi}=(V,J)$ is the steady state solution given by
(\ref{10.72}), and $\psi_R^\pm$  are  the stable equilibria of
(\ref{10.73}), derived in Theorem~\ref{t10.2}.  
In fact, the ENSO phenomenon can be explained as the transition
between the two metastable steady state solutions $\tilde{\psi}+\psi_R^\pm$.

 Theorem \ref{t10.2}  amounts to saying that for a mixed
transition, there are two critical Rayleigh numbers $R^*$ and
$R^{\varepsilon}_c$, with
$R^* < R^{\varepsilon}_c$,  where $R^\ast \cong R_c\cong 2.7\times
10^{21}$. 
At $R=R^*$, the equation (\ref{10.73}) has a saddle-node
bifurcation, and at $R=R^{\varepsilon}_c$ has the normal transition,
as shown in Figure \ref{f10.10}. When the Rayleigh number $R$ is less than
$R^*$, i.e. $R<R^*$, the system (\ref{10.39}) has only one stable
steady state solution $\tilde{\psi}=(V,J)$ as in (\ref{10.72}) which
attracts $H$. When $R^*<R<R^{\varepsilon}_c$, this system has two
stable equilibriums
\begin{equation}\tilde{\psi}\in U^R_+,\ \ \ \ \text{and}\ \ \ \
\tilde{\psi}+\psi^-_R\in U^R_-,\ \ \ \ \text{for}\ \ \ \
R^*<R<R^{\varepsilon}_c,\label{10.79}
\end{equation}
where $U^R_+$,  $U^R_-\subset H$ are basins of attraction  of $\tilde{\psi}$
and $\tilde{\psi}+\psi^-_R$. And when $R^{\varepsilon}_c<R$, this
system has also two stable equilibriums
\begin{equation}
\tilde{\psi}+\psi^-_R\in U^R_-\ \ \ \ \text{and}\ \ \ \
\tilde{\psi}+\psi^+_R\in U^R_+ \ \ \ \ \text{for}\
R^*_c <R,\label{10.80}
\end{equation}
with $U^R_-$ and $U^R_+$ are their basins of attraction. 
\begin{figure}[hbt]
  \centering
 \includegraphics[width=0.7\textwidth]{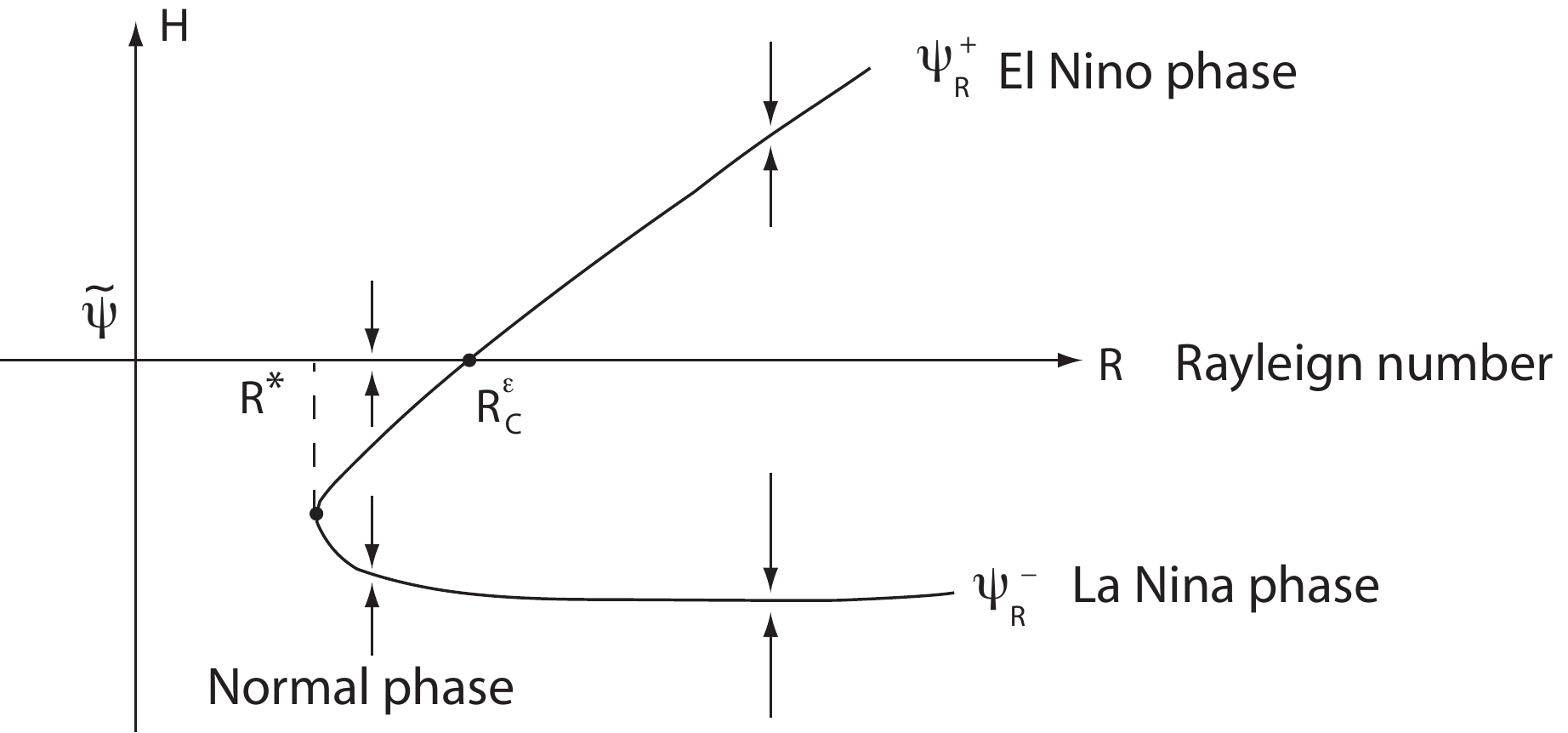}
  \caption{The transition diagram for the equatorial motion
equations (\ref{10.39}), $R^*$ is the saddle-node bifurcation point,
$R^{\varepsilon}_c$ is the first critical Rayleigh number.}\la{f10.10}
 \end{figure}

Since the problem (\ref{10.39}) with a natural boundary condition is
a perturbation for the idealized condition, 
$\tilde \psi=(V, J)$  is small, and the velocity field $V$ is almost zero. Hence the equilibrium 
$$\Psi^+_R = \left\{
\begin{aligned}
& \tilde \psi   && \tf R < R^\ast_c, \\
&  \tilde \psi + \psi^+_R   && \tf R > R^\ast_c, 
\end{aligned}
\right.
$$
represents the El Ni\~no phase, and the equilibrium 
$$\Psi^-_R =   \tilde \psi + \psi^-_R   
$$
represents the normal phase   for $R^\ast < R < R^\ast_c$, and the La Ni\~na phase for 
$R^\ast_c < R$.  
It is clear that both  $\Psi^\pm_R$ are metastable, with $U^\pm_R$  as their basins of attractions respectively.

\bigskip
{\bf Oscillation Mechanism of ENSO.}
The above theoretical studies suggest that  ENSO is the {\it interplay} between two  oscillation processes. The first is the oscillation between the metastable  warm event (El Ni\~no phase, represented by $\Psi^+_R$)  and cold event (La Ni\~na phase, represented by $\Psi_R^-$). The second oscillation is the oscillation of the Rayleigh number caused essentially by the some spatiotemporal oscillation of the  sea surface temperature (SST) field.

 Here we present  a brief schematic description on the interplay between these two oscillation processes of the ENSO, based on the saddle-node transition diagram in Figure~\ref{f10.10}, rigorously proved in Theorem~\ref{t10.2}.

We start with three physical conclusions:

\begin{itemize}

\item[1] We observe that  as $R$ decreases, the normal and the La Ni\~na phase  $\Psi^-_R$ weakens  and its basin of attraction shrinks (to zero as $R$ approaches to $R^\ast$). 

\item[2] As $R$ increases in the interval $R^\varepsilon_c< R$, the strength of  the El Ni\~no phase increases. 
As $R$ increases in $(R^\ast, R^\varepsilon_c)$, the basin of attraction of the El Ni\~no phase shrinks. In particular, near 
$R^\varepsilon_c$, the El Ni\~no phase is close to the interaction of the two basins of attraction of the El Ni\~no and La Ni\~na phase; consequently, forcing the transition from the El Ni\~no phase to the La Ni\~na phase.

\item[3] Also, we see that as the El Ni\~no event  $\Psi^+_R$ intensifies,  the SST increases,   and as the normal and La Ni\~na event   $\Psi^-_R$ intensifies, the SST decreases. 

\end{itemize}

From the above three physical conclusions, we obtain the following new mechanism of the ENSO oscillation process: 
\begin{itemize}

\item[I] When $R^\ast < R < R^\varepsilon_c$,  $\Psi^-_R$  represents the normal condition, and the corresponding upwelling near Peru leads to the decreasing of the SST, and leads to $R$ approaching $R^\ast$. 
On the other hand, near $R^\ast$, the basin of attraction of the normal condition $\Psi^-_R$ shrinks, and due to the uncertainty of the  initial data,  the system can undergo a dynamic transition near $R^\ast$ toward to the El Ni\~no phase $\Psi^+_R$.

\item[II] For the El Ni\~no phase $\Psi^+_R= \tilde \psi$ near $R^\ast$, however, the velocity $V$ is small, and SST will increases due to the solar heating, which  can not be  transported away  by weak oceanic currents without upwelling. Hence the corresponding Rayleigh number $R$ increases, and the El Ni\~no phase intensifies.
When the Rayleigh number approaches to the critical value $R^\varepsilon_c$, with high probability, the system undergoes a metastable transition from  the El Ni\~no phase to either the normal phase or the La Ni\~na phase depending on the Rayleigh number $R$ and the the strength of the phase $\Psi^-_R$.

\item[III] With delaying effect,  strong El Ni\~no and La Ni\~na occur at  the Rayleigh number  $R$ larger than the critical 
number $R^\varepsilon_c$, and the system will repeat  the process described in items I-II above.\end{itemize}

In summary, this new  mechanism of ENSO as an interplay of the two processes  leads to both the random and deterministic features of the ENSO, and defines a new natural {\sc feedback} mechanism, which drives the sporadic oscillation of the ENSO. The randomness is closely related to the uncertainty/fluctuations of the initial data between the narrow basins of attractions of the corresponding metastable events, and the deterministic feature is represented by a deterministic coupled atmospheric and oceanic model predicting the basins of attraction and the SST. It is hoped this mechanism based on a rigorous mathematical theory could lead to a better understanding and prediction of the ENSO phenomena.

\bibliographystyle{siam}

\end{document}